\documentclass[12pt]{JHEP3}

\let\ifpdf\relax
\usepackage{ifpdf}
\usepackage{graphics}
\usepackage{graphicx}
\usepackage{amssymb}
\usepackage{amsmath}
\usepackage{slashed}
\usepackage{cite}
\usepackage{epsfig}
\usepackage{datetime}
\usepackage{enumerate}

\newcommand{\be}{\begin{equation}}
\newcommand{\ee}{\end{equation}}
\newcommand{\bal}{\begin{align}}
\newcommand{\eal}{\end{align}}
\newcommand{\bea}{\begin{eqnarray}}
\newcommand{\eea}{\end{eqnarray}}

\newcommand{\bi}{\begin{itemize}}
\newcommand{\ei}{\end{itemize}}

\def\RR{{\mathbb R}}

\def\tr{{\rm tr}}

\title{Mutual R\'enyi information for two disjoint compound systems}

\author{Howard J. Schnitzer  \\
{Martin Fisher School of Physics, Brandeis University, \\ \ \ \ \ \ Waltham, MA 02454, USA\\
}
\email{Schnitzr@brandeis.edu}}

\preprint{
 BRX-TH-679}

\abstract{
The leading term for the mutual R\'enyi  information is studied for two widely separated identical compound systems for free scalar fields in $(d+1)$ Euclidean space. The compound system consists of two identical spheres in contact, with a result consistent with a universal form for the leading term for the mutual R\'enyi  information.
}

\begin{document}

\section{Introduction}
$\,$

The study of entanglement entropy has become a significant quantity for the understanding of the global structures of a quantum field theory. This involves the von Neumann entropy
\bea
S_A:=-\tr(\rho_A\ln \rho_A)\,,
\eea
computed from the reduced density matrix $\rho_A$ of a subsystem $A$ obtained from the density matrix $\rho$ of the theory by tracing over the complementary subsystem. In this connection one is interested in the $n^{th}$ R\'enyi entanglement entropy defined as 
\bea
S_A^{(n)}=(n-1)^{-1}\ln\tr(\rho_A^n)
\eea
with 
\bea
S_A=\lim_{n\to 1} S_A^{(n)}
\eea
where explicit computations of $S_A^{(n)}$ involve the so called replica trick \cite{Holzhey:1994we,Calabrese:2004eu}. Another quantity of interest is the mutual information of two disjoint compact regions $(A,B)$ obtained from the $n\to 1$ limit of the R\'enyi entropies \cite{Calabrese:2009ez}
\bea
I^{(n)}(A,B)=S_A^{(n)}+S_B^{(n)}-S_{A\cup B}^{(n)}
\eea

Consider a free scalar field in $\RR^{d+1}$ with action 
\bea
\label{action}
\frac 12 \int d^{d+1}(\partial \phi)^2
\eea
Cardy \cite{Cardy:2013nua} considers the mutual R\'enyi information entropy for two disjoint regions $(A,B)$ given by 
\bea
\label{mutualdis}
I^{(n)}(A,B)=(n-1)^{-1}\ln\left[\frac{Z(C^{(n)}_{A\cup B}) Z^{(n)}}{Z(C^{(n)}_{A})Z(C^{(n)}_B)}\right]\,,
\eea
where $Z(C^{(n)}_{X})$ is the partition function on the conifold $C_{X}^{(n)}$ which is a $d-1$ dimensional sub-manifold of conical singularities along the boundary $\partial X \cap \{\tau =0\},$ and where $\RR^{d+1}=\RR^d\times \tau,$ with $\vec{r}\in \RR^{d}$, and $X\in (\vec{r},\tau)$, with $\tau$ thought of as Euclidean time. Equation (\ref{mutualdis}) is interesting as it is expected to only depend on the geometry of $A$ and $B$, and the data of the quantum field theory (QFT). Various examples give new insights, particularly for entanglement across complicated boundaries. It is the purpose of this paper to treat one such example, particularly since mutual information for a free scalar field in higher dimensions has been studied in only few papers \cite{Casini:2009sr,Shiba:2010dy,Shiba:2012np}.

For $A$ and $B$ asymptotically separated one can consider $A$ and $B $ separately to obtain the leading term in the large separation $r$ \cite{Cardy:2013nua}
 \bea
 \label{leadmutualinf1}
 I^{(n)}(A,B)\sim g_d^{(n)}\left( \frac {R_A R_{B}}{r^2}\right)^{d-1}
 \eea
 or an alternate form, based on our results 
 \bea
 \label{leadmutualinf2}
 I^{(n)}\sim  \frac {h_d^{(n)} C_A C_{B}}{r^{2 (d-1)}}
 \eea
 where $C_A$ and $C_B$ are the electrostatic capacities of $A$ and $B$, $R_A$ and $R_B$ are their linear sizes, and $g_d^{(n)}$ and $h_d^{(n)}$ are dimensionless numbers which characterize the particular system. In this paper we consider the QFT (\ref{action}) and with $A$ and $B$ each two identical spheres in contact. [Four spheres in all]. We find 
\bea
\label{CA}
 C_A=\frac{(2R_A)^{d-1}}{2^{d-2}}\sum_{l=1}^{\infty}\frac{(-1)^{l+1}}{l^{d-1}}
\eea
where $R_A$ is the radius of one of the spheres in $A$, and similarly for $B$. In particular we obtain 
\bea
I^{(n)}(A,B)&\sim& \frac{2n}{r^{2(d-1)}}(R_A R_B)^{d-1}\left[\sum_{l=1}^{\infty}\frac{(-1)^{l+1}}{l^{d-1}}\right]^2 \nonumber
\\ 
&& \left\{\frac14 \sum_{j=1}^{n-1}(\frac 1j)^2 +\frac{(n-1)^2}{n^2}\right\}(n-1)^{-1} \label{Infor}  \,.
\eea
 
In our calculations we need the free Green's function appropriate to (\ref{action}) i.e.
\bea
\langle \phi(x_1)\phi(x_2)\rangle&=& G_0(x_1-x_2)\nonumber \\
&=&|x_1-x_2|^{-(d-1)}
\eea
We also need 
\bea
\label{normalorder}
:\phi^2(x):=\lim_{\delta \to 0}\Big[\phi(x+{\delta}/{2})\phi(x-{\delta}/{2})-G_0(\delta)\Big]
\eea
and
\bea
\langle:\phi^2(x_1)::\phi^2(x_2):\rangle&=& 2G_0(x_1-x_2)^2 \,.
\eea
The conifold satisfies the boundary conditions \cite{Calabrese:2004eu,Calabrese:2009ez,Agon:2013iva}
\bea
\phi_j(r,0_-)&=&\phi_{j+1}(r,0_+)\qquad\textrm{    for $r\in A$}
\nonumber \\
&=&\phi_{j}(r,0_+)\qquad    \qquad r\notin A
\eea
and similarly for $B$ separately when $A$ and $B$ are widely separated, as in (\ref{leadmutualinf1}-\ref{leadmutualinf2})
[In sec.2 ff, we find it convenient to use dimensionless Green's functions and adjust the normalization at the end of the calculation].

Cardy \cite{Cardy:2013nua} shows that the leading contribution to (\ref{leadmutualinf1}-\ref{leadmutualinf2}) is obtained from his equation (8), i.e.
\bea
\frac12 r^{-2(d-1)}n\Bigg[\sum_{j=1}^{n-1}\left(\lim_{x_1\to \infty}(x_1^2)^{d-1}\langle \phi_j(x_1)\phi_0(x_1)\rangle_{C^{(n)}_A}\right)\left(\lim_{x_1\to \infty}(x_1^2)^{d-1}\langle \phi_j(x_1)\phi_0(x_1)\rangle_{C^{(n)}_B}\right)\nonumber \\ + \left(\lim_{x_1\to \infty}(x_1^2)^{d-1}\langle :\phi_0(x_1)^2:\rangle_{C^{(n)}_A}\right)\left(\lim_{x_1\to \infty}(x_1^2)^{d-1}\langle :\phi_0(x_1)^2:\rangle_{C^{(n)}_B}\right) \Bigg]\nonumber\\
\eea
Further, the (conformal) inversion symmetry maps correlation functions on $C_A^{(n)}$ to  $C_A^{'(n)}$ which requires Green's functions obtained from the inversion map. In particular the inversion maps the double spheres of $A$ to two parallel hyperplanes. Both $A$ and $B$ can be described after inversion as \bea  
\label{2dcone}
[\{\textrm{2-dimensional cone: opening angle}\quad 2\pi k \} \times \RR^{d-1}]
\eea

The structure of the paper is as follows: In section 2 we compute the capacity of the two spheres and Green's functions for $k=1$, in equations (\ref{qt}, \ref{greenn}). In section 3, we compute the mutual R\'enyi information using the Green's functions appropriate to (\ref{2dcone}) for $k\ge 1$. In section 4 we summarize our results and offer some conclusions.

\section{\label{2} Two identical spheres}
\subsection{Capacity}
${\,}$

As a preliminary for the calculation of the mutual information for the two spherical pairs, we use the inversion map to compute the capacity of two identical spheres in contact in $\RR^D=\RR^{d+1}\,.$ To do so, we make use of the inversion map in $D$-dimensions, as follows. Let $\phi(r)$ be a solution of Laplace's equation in $D$-dimensions. Consider the inversion through the origin by a sphere of radius $a$. Then, there is another solution of Laplace's equation
\bea
\phi^I(\vec{r}\,)&=&\left(\frac{a}{r}\right)^{D-2}\phi\left(\frac{a^2}{r}\vec{r}\right)\,, \nonumber \\  
&=&\left(\frac{a}{r}\right)^{D-2}\phi(\vec{r}^{\,\, I})\,.
\eea
Use the inversion map of the two spheres at potential $V$ in contact in $\RR^D$, with center of inversion the point of contact of the two spheres. The inversion radius $a$ is the diameter of one of the spheres. The two spheres map to two parallel grounded planes separated by $2a$, after subtracting potential $V$ from the original system. 
There are an infinite set of image charges
\bea
q'_n=(-1)^n a^{D-2}V
\eea
beyond the grounded planes at a distance 
\bea
\label{rn}
r'_n=(2n+2) a
\eea
from the center of the inversion sphere. 
Map back to obtain the image charges for the original two spheres, which are 
\bea
\label{charges}
q_n&=&q'_n\left(\frac{a}{r'_n}\right)^{D-2} \nonumber \\
&=&\frac{(-1)^n\,a^{D-2}V}{[2(n+1)]^{D-2}}
\eea
inside each sphere. The total charge stored on the two spheres is 
\bea
\label{qt}
q_{tot}&=&2\sum_{n=0}^{\infty} q_n\nonumber \\ 
&=&\frac{a^{D-2}}{2^{D-3} } \sum_{n=1}^{\infty} \frac{(-1)^{n+1} V}{n^{D-2}}
\eea
so that the capacity for the two spheres is 
\bea
C_A&=&\frac{a^{D-2}}{2^{D-3}} \sum_{n=1}^{\infty} \frac{(-1)^{n-1} }{n^{D-2}}\,, \nonumber \\ 
&=&\frac{a^{d-1}}{2^{d-2}} \sum_{n=1}^{\infty} \frac{(-1)^{n+1}}{n^{d-1}}   
\eea
which reduces to the familiar 
\bea
C_A=2R_A\ln 2
\eea
for $D=3$ with $a=2R_A.$\\
\subsection{Green's function}
$\,$

The contribution of the $n^{th}$ image charge to the Green's function for this system (before inversion) is
\bea
\label{greenn}
G_n(\vec{r},\vec{r}_n)&=&\frac{q_n}{|\vec{r}-\vec{r}_n|^{D-2}} \nonumber \\
&=& \frac{q_n}{[r^2+r^2_n-2\,\vec{r}\cdot\vec{r}_n]^{(D-2)/2}} \nonumber \\
&=& \frac{q_n}{\{ r ^2+\frac{a^2}{4(n+1)^2}-\frac{r\,a}{(n+1)}\hat{r}\cdot\hat{r}_n \}^{(D-2)/2}} 
\eea
where $q_n$ is given by (\ref{charges}) and $|\vec{r}_n|$ for the position of the image charge is 
\bea
r_n=\frac{a^2}{r'_n}=\frac{a}{2(n+1)}
\eea
satisfying $r_nr'_n=a^2$. The angle between an arbitrary observer at $\vec{r}$ and the position of the charge at $\vec{r}_n$ is
\bea
\cos \gamma_n=\hat{r}\cdot \hat{r}_n
\eea

The $z$-axis lies along the common diameter of the two spheres, with origin at the point of contact. Therefore
\bea
\vec{r}_n=\pm r_n \hat{z}
\eea
locates the image charges in the original two sphere configuration. 

In section 3, we generalize the strategy of this section so as to compute the mutual R\'enyi information for a pair of two-spheres, widely separated using the inversion map and method of images.

\section{Mutual R\'enyi information}
\subsection{The inversion map}
$\,$

Consider the free field Lagrangian (\ref{action}) in $\RR^{d+1}$ with $A$, a pair of d-dimensional identical balls of radius $a/2$ in contact, and similarly for $B$. The common equatorial plane of the system is at $\tau=0$, where $\tau$ is the Euclidean time. This intersection yields two pairs of two-spheres in contact. That is, the intersection is 
\bea
\{S^{d-1}\times S^{d-1} \}\oplus\{S^{d-1}\times S^{d-1} \}\,,
\eea
which describes the two subsystem $C_A$ and $C_B$. Consider $C_A$ and $C_B$ to be widely separated, so that $C_A$ and $C_B$ can be analyzed individually. We proceed to do so. 

Consider the inversion of $A$, as in section 2, which maps the two balls to two hyperplanes in $\RR^d$. The plane $\tau=0$ is preserved in the map, so that $\{S^{d-1}\times S^{d-1} \}$ is mapped to two parallel hyperplanes $\RR^{d-1}$. That is, $\partial A \cap \{\tau=0\}$ is mapped to two parallel hyperplanes $\RR^{d-1}$ separated by $2a$.

We generalize this set-up so as to compute the mutual R\'enyi information. Before carrying out an inversion, consider the conifold obtained by rotating by an angle $\theta$ about the common $z$-axis of the pair of balls, with $\theta$ identified by $\theta+2\pi k$. Under the inversion map, each sphere $S^{d-1}$ of the pair in $A$ is mapped to 
\bea
\label{consing}
\{\textrm{2-dimensional conical singularity} \}\times \RR^{d-1} \,,
\eea
so that the system $A$ is mapped under inversion to
\bea
\label{2dcone}
[\{\textrm{2-dimensional cone} \}\times \RR^{d-1} ]_{left}\oplus [\{\textrm{2-dimensional cone} \}\times \RR^{d-1} ]_{right}\,,
\eea
and similarly for $B$. That is (\ref{consing}) is the inversion for singularities along $\partial A\cap\{\tau=0\}$. Since $A$ and $B$ are asymptotically separated, we consider the inversion of $A$ and $B$ individually. It is important to note that the ``left-cone '' and ``right-cone'' in (\ref{2dcone}) have the same opening angle, since the cones arise from a single rotation about the common $z$-axis of the two spheres in contact. That is, each sphere of $C_{A}^{(k)}$ is mapped to a hyperplane of conical singularities, as in (\ref{consing}). 
Denote that system $C_A^{'(k)}$, so that $C_A^{(k)}$ and $C_A^{'(k)}$ are related by the inversion map, and similarly $C_B^{(k)}\to C_B^{'(k)}$. We have the same conifold angle for $A$ and $B$ if the entire system is rotated about the common $z$-axis of the original configuration. 

\subsection{Green's function: $k \geq 1$}
$\,$

In this section we generalize the strategy of Section 2.2, to include the consequence of the conifolds. Use cylindrical coordinates for $C_{A}^{'(k)}$, i.e. $(\rho,\theta,\bar{z})$ where $\theta \in [0, 2\pi k]$ and $\vec{z} \in \RR^{d-1}$ for each hyperplane of $C_A^{'(k)}$, which recall are separated by $2a$. We write the Green's function for $C_A^{'(k)}$ schematically 
\bea
\label{green}
G^{'(1)}=G^{'(1)}_L+G_R^{'(1)}\,,
\eea
which are due to the image charges of the left and right hyperplanes respectively. 

For a general observer and $k=1$
\bea
\label{greenr}
G_R^{'(1)}=\sum_{n=0}^{\infty}\frac{(-1)^n a^{d-1}}{\{  \rho^2 +[(2n+2)a]^2 -2\rho(2n+2)a\cos \theta \}^{(d-1)/2}}\,,
\eea
since the image charges are all at $\vec{z}=0$, and similarly for $G^{'(1)}_L$. However in our problem, $\vec{r} \to \infty$ in $C_A^{(k)}$ is mapped to $\vec{r}{\,'}=0 \in C_A^{'(k)}$ which implies $\rho\to 0$. Therefore, from (\ref{greenr})
\bea
\label{greenr2}
G_R^{'(1)}\xrightarrow[\rho\to 0]{} \frac{1}{2^{d-1}} \sum_{n=0}^{\infty}\frac{(-1)^n }{(n+1)^{d-1}}\,,
\eea
[Note that our Green's functions are dimensionless. At the end of our calculations we change the normalization to obtain the R\'enyi information].

Now consider the Green's function (\ref{green}) for $k\geq 1$, which satisfies the periodicity condition
\bea
\label{greenk}
G^{'(k)}(\rho,\theta +2\pi k,\vec{z} )=G^{'(k)}(\rho,\theta,\vec{z})\,,
\eea
for the two hyperplanes of (\ref{2dcone}), i.e. for $C_A^{'(k)}$. Following Cardy's idea \cite{Cardy:2013nua} , we analytically continue (\ref{greenk}) to $k=1/m,$ where $m$ is a positive integer. Then by the method of images, each charge $q_n^{'}$ gives rise to $m$ images, such that (\ref{greenk}) is satisfied. Thus (\ref{greenr}) and (\ref{greenr2}) generalize to 
\bea
G_R^{'(1/m)}&=&\lim_{\rho\to 0}\sum_{j=0}^{m-1}\sum_{l=0}^{\infty}\frac{(-1)^l a^{d-1}}{\{  \rho^2 +[(2l+2)a]^2 -2\rho \cos (\theta+2{\pi j}/{m} )\}^{(d-1)/2}}\,, \nonumber \\
&=& m\,G_R^{'(1)} \,, \\
&=&\frac{m}{2^{d-1}}\sum_{l=1}^{\infty}\frac{(-1)^{l+1}}{l^{d-1}} \,, \nonumber
\eea
Thus, from (\ref{greenr2}) and $m=1/k$
\bea
\label{greenk1}
G^{'(k)}&=&\frac{m}{2^{d-2}}\sum_{l=1}^{\infty} \frac{(-1)^{l+1}}{l
^{d-1}}\,, \\ \label{greenk2}
&=& \frac 1k\frac{1}{2^{d-2}}\sum_{l=1}^{\infty} \frac{(-1)^{l+1}}{l
^{d-1}} \,.
\eea

\subsection{R\'enyi information}
$\,$

In order to compute the R\'enyi mutual information we need from (\ref{greenk1}-\ref{greenk2})
\bea
\sum_{j=1}^{n-1}[\,G^{'(j)} \,]^2=\frac{16}{2^{2d}}\left[\sum_{l=1}^{\infty}\frac{(-1)^{l+1}}{l^{d-1}}\right]^2 \sum_{j=1}^{n-1} \left(\frac 1j\right)^2 \,,
\eea
where
\bea
\sum_{j=1}^{n-1} \left(\frac 1j\right)^2=H_{n-1}^{(2)}\,,
\eea
is the generalized Harmonic number. Further we need to compute $:\phi^{2}(x):$ using (\ref{normalorder}). That is
\bea
[\,G^{'(n)}-G^{'(1)}\,]=\frac{(1-n)}{n}\frac{1}{2^{d-1}}\sum_{l=1}^{\infty}\frac{(-1)^l}{l^{d-1}}\,.
\eea
Finally, in order to make use of Cardy's equations (7) and (8) we need to 
\begin{enumerate}[i]
\item Change the dimension of the Green's functions\,,
\item Map $G^{'(j)}$ to $G^{(j)}$, which is obtained from the map of $C_A^{'(j)}$ to $C_A^{(j)}$.
\end{enumerate}

 Putting this all together, we obtain (\ref{Infor}) for the leading contribution 
 \bea
 I^{(n)}(A,B)\sim \frac{n}{2r^{2(d-1)}} C_{A} C_{B}\left\{ \frac 14 \sum_{j=1}^{n-1}\left(\frac 1j\right)^2 +\frac{(n-1)^2}{n^2} \right\}(n-1)^{-1}\,,
 \eea
 where $a=2R_A=2R_B,$ and $C$ is as in (\ref{CA}). As a check, we find for $n=2$
 \bea
 I^{(2)}(A,B)\sim \frac 1{2 r^{2(d-1)}}C_A
 C_B\,,
 \eea
 in agreement with Cardy, equation (11).
\section{Conclusions}

We have computed the leading term for the R\'enyi mutual information of two disjoint regions of a pair of double spheres asymptotically separated. The result (\ref{leadmutualinf1}-\ref{leadmutualinf2}) has the same general form of a pair of single spheres widely separated, so that (\ref{leadmutualinf1}-\ref{leadmutualinf2}) may well be universal in form for free scalar fields in $\RR^{d+1}$, with only the overall numerical coefficient depending on the system in question. The behavior for finite separations and various systems, and generalizations of (\ref{action}) are more difficult problems which deserve attention. It would also be interesting to understand to what extent this class of problems can be described holographically \cite{Headrick:2010zt,Ryu:2006bv,Ryu:2006ef,Tonni:2010pv}. \textbf{Note added:} Relevant results have appeared since this paper was originally posted \cite{Herzog:2014fra}, \cite{Shiba:2014uia}.

\paragraph{Acknowledgments: }
The calculation presented in this paper was strongly influenced by the analysis of Cardy. We thank him for reading the manuscript. We also wish to thank Matthew Headrick and Isaac Cohen for conversations, and C\'esar Ag\'on for his invaluable assistance in preparing this manuscript. HJS is supported in part by the DOE by grant DE-SC0009987.

\eject
\bibliographystyle{utphys}
\bibliography{rmirefs}

\end{document}